# Adaptive Vision-Based Coverage Optimization in Mobile Wireless Sensor Networks: A Multi-Agent Deep Reinforcement Learning Approach

P. Soltani, M. Eskandarpour, S. Heidari, F. Alizadeh, H. Soleimani
School of Electrical Engineering, Iran University of Science & Technology, Tehran, Iran

*Abstract*— Traditional Wireless Sensor Networks (WSNs) typically rely on pre-analysis of the target area, network size, and sensor coverage to determine initial deployment. This often results in significant overlap to ensure continued network operation despite sensor energy depletion. With the emergence of Mobile Wireless Sensor Networks (MWSNs), issues such as sensor failure and static coverage limitations can be more effectively addressed through mobility. This paper proposes a novel deployment strategy in which mobile sensors autonomously position themselves to maximize area coverage, eliminating the need for predefined policies. A live camera system, combined with deep reinforcement learning (DRL), monitors the network by detecting sensor LED indicators and evaluating real-time coverage. Rewards based on coverage efficiency and sensor movement are computed at each learning step and shared across the network through a Multi-Agent Reinforcement Learning (MARL) framework, enabling decentralized, cooperative sensor control. Key contributions include a vision-based, low-cost coverage evaluation method; a scalable MARL-DRL framework for autonomous deployment; and a self-reconfigurable system that adjusts sensor positioning in response to energy depletion. Compared to traditional distance-based localization, the proposed method achieves a 26.5% improvement in coverage, a 32% reduction in energy consumption, and a 22% decrease in redundancy, extending network lifetime by 45%. This approach significantly enhances adaptability, energy efficiency, and robustness in MWSNs, offering a practical deployment solution within the IoT framework.

## I. INTRODUCTION

Wireless Sensor Networks (WSNs) are becoming an essential platform for environmental monitoring [1], infrastructure inspection [2], industrial automation [3], and other mission-critical applications due to distributed sensing and data collection across large physical areas [4]. Sensor nodes in WSNs are traditionally deployed statically, with pre-selected deployment locations based on the target area, expected coverage demands, and network constraints. To ensure network reliability and redundancy, static WSNs usually involve significant sensor overlap to maintain coverage even if individual nodes deplete their energy storage or fail [5]. However, the static nature of this strategy makes it rigid, especially in dynamic or uncertain environments where coverage requirements can change or node failures are common [6].

Mobile Wireless Sensor Networks (MWSNs) improve on the benefits of conventional WSNs by allowing sensor nodes to move around, resulting in greater coverage flexibility, adaptive recovery behavior, and fault tolerance [7]. Mobile sensors can travel to uncovered areas, adapt to dynamic environments, and replace faulty nodes by repositioning themselves through controlled movement—helping to maintain network functionality over extended periods despite energy constraints [8]. Mobility presents challenges, particularly in autonomous deployment, real-time localization, adaptive coordination, and power efficiency. In such cases, dynamic deployment and localization are critical for maintaining meaningful sensing and communication in MWSNs [9]. Unlike static nodes that are deployed in fixed positions, mobile sensors, require recurrent localization to accurately process sensed data, maintain network topology, and enable cooperative decision-making [10].

Conventional localization techniques, such as GPS and range-based techniques (e.g., RSSI, TOA, TDOA), while beneficial in some cases, have drawbacks such as high power consumption, high cost, susceptibility to environmental noise, and lack of availability in GPS-denied environments such as indoors, underground, or underwater [11]. Range-free localization techniques require less energy but are typically less precise. Furthermore, pre-configured deployment plans cannot be easily scaled up in dynamic environments where sensor positions and coverage requirements change over time. In this paper, we propose a novel DRL-based autonomous deployment system for MWSNs that allows mobile sensors to self-organize and maximize their placement for optimal area coverage. Our system eliminates pre-defined deployment schemes, allowing the network to adapt to dynamic environments and node failures in real time. One of the distinguishing features of our work is the use of a live camera system installed in the environment to collect visual observations—LED indicators on sensor nodes—for the purpose of analyzing coverage quality. Visual information is processed using computer vision techniques and translated into reward signals based on sensor placement performance.

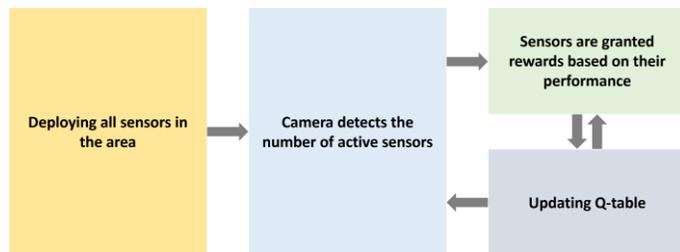

**Fig. 1**. Deep Reinforcement Learning steps.

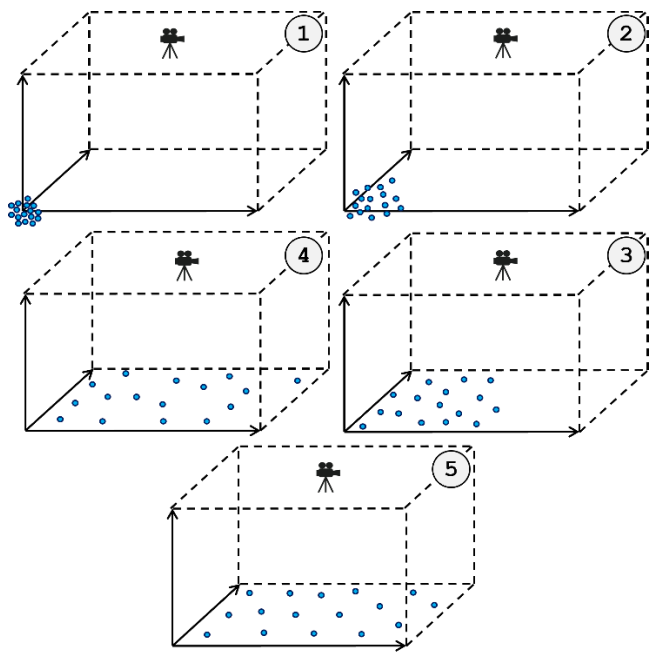

**Fig. 2**. Localization of the sensors in MWSN.

For vision-based sensor identification and localization, a camera is mounted above the deployment area and captures images every 10 seconds during training and evaluation. Each sensor is equipped with a red LED indicator to signify its active state. We employ a convolutional neural network (CNN) model trained to detect and localize active sensors by identifying the distinct LED signatures in the captured images. The CNN architecture consists of three convolutional layers with 3×3 kernels, each followed by ReLU activation and max pooling, and two fully connected layers that output the pixel coordinates of detected LEDs. The model was trained on a synthetic dataset of labeled sensor images generated under varied lighting and angle conditions. The system achieves an average localization accuracy of 95.2% on a test set, measured by intersection-over-union (IoU) between predicted and actual LED locations. To ensure robustness against occlusion and environmental noise, image pre-processing includes Gaussian blurring and adaptive thresholding, followed by bounding box extraction. In cases of partial occlusion or LED fading, temporal filtering is used to verify LED consistency across consecutive frames. To handle variations in lighting or reflections, histogram equalization is applied to normalize brightness before CNN inference. Additionally, to correct for minor distortions due to camera misalignment or lens distortion, we perform geometric calibration using a planar homography derived from known marker positions on the deployment field. These combined strategies make the vision-based detection pipeline robust and reliable across a range of real-world disturbances, with minimal degradation observed during moderate occlusion, non-uniform lighting, or slight shifts in camera orientation.

These rewards are then applied to a multi-agent deep reinforcement learning (MADRL) architecture, in which each sensor learns its own movement strategy independent of the global coverage outcome [12]. Such a vision-based and decentralized scheme provides several significant benefits. First, it can use GPS-free localization with visual feedback, making it suitable for deployment in GPS-denied environments [13]. Second, the use of MADRL, which is less centralized in computation or communication, facilitates distributed learning and coordination among sensors. Third, dynamic reconfiguration of the system enables the network to dynamically relocate sensors in response to energy depletion, environmental obstacles, or changing coverage requirements [14].

Finally, energy-conscious learning prevents unnecessary relocation and extends the network's lifetime. To address these limitations, recent research has investigated the use of machine learning, and more recently, deep reinforcement learning (DRL), to manage sensor networks [15]. DRL allows agents to learn their optimal actions by interacting with the environment, rather than relying on explicit models or predefined policies. In MWSNs, each mobile sensor can be modeled as an autonomous agent capable of sensing its surroundings, receiving feedback on its actions, and learning to change its location in order to optimize global network performance. Unlike traditional control methods, DRL excels at dealing with the high-dimensional, non-linear, and partially observable nature of mobile sensor deployments, especially in real-world, uncertain environments.

## II. RECENT STUDIES

he field of Mobile Wireless Sensor Networks (MWSNs) has evolved significantly over the past two decades, particularly in the area of localization and deployment strategies. Early studies primarily focused on signal-based localization techniques to determine sensor positions and maintain network connectivity. Amundson et al. [16] proposed a method that utilizes RF Doppler shifts for sensor localization in mobile environments.

Their approach was particularly effective in GPS-denied scenarios, offering improved navigation accuracy by analyzing frequency variations in received signals. Similarly, Fang et al. [17] developed NavMote, a pedestrian dead reckoning system that combines inertial sensor data with wireless signals to track user movement. This method enhances position estimation accuracy in environments where GPS signals are unreliable or unavailable, such as indoors or underground facilities.

Efforts to enhance scalability and energy efficiency have led to innovative solutions like ZebraNet, introduced by Juang et al. [18]. This system utilized energy-efficient peer-to-peer networking in wildlife tracking, leveraging MWSNs to monitor zebra movements in large, remote areas. The study showed how distributed computing and mobility could extend network coverage without relying on fixed infrastructure. Kusy et al. [19] further built on RF Doppler principles by tracking mobile nodes through motion-induced frequency changes, demonstrating how mobile nodes can localize themselves without external infrastructure or GPS dependency.

Several research efforts have incorporated mobility as a core component of deployment and localization. Dantu et al. [20] developed RoboMote, a mobile sensor platform that allows sensors to autonomously reposition themselves to enhance coverage. Similarly, Ragobot by Friedman et al. [21] emphasized self-localization and repositioning for improved adaptability and network efficiency. Both systems highlighted the importance of robotic mobility, but their reliance on centralized or predefined movement strategies limited their flexibility in dynamic environments. Bergbreiter and Pister [22] presented CotsBots, a low-cost distributed robotics platform designed for MWSNs. Their work demonstrated that off-the-shelf mobile robotic systems could be integrated into wireless networks for autonomous sensing and navigation, offering affordability and accessibility. However, most of these systems still depended on idealized simulation models and did not account for real-time sensor failures or energy depletion.

More recent work has focused on resilience and fault tolerance in sensor networks. Wang et al. [23] proposed a sensor relocation method that allows mobile nodes to replace failed static nodes, preserving network coverage in the face of hardware failures. Liu et al. [24] explored how sensor mobility could dynamically improve coverage and localization by responding to environmental changes. Gandham et al. [25] addressed the issue of power consumption in MWSNs by optimizing the movement of multiple mobile base stations, showing that energy-aware repositioning can reduce communication overhead and extend network lifespan. These studies illustrate the ongoing shift from static, signal-reliant strategies toward more flexible, energy-efficient approaches based on mobility and adaptive reconfiguration.

Despite these advances, several common limitations remain. Methods based on signal strength measurements or Doppler shifts [16–20] often require high energy for communication and are susceptible to environmental interference. Moreover, systems like RoboMote [20] and Ragobot [21], while supporting mobility, operate under centralized control or pre-programmed movement policies, which limits their scalability and real-time adaptability in dynamic or unknown environments. Even studies investigating autonomous mobility [22–25] tend to rely on idealized assumptions—such as perfect sensing models or error-free communication—which can compromise performance in real-world deployments. Most notably, few of these methods address how networks adapt in real time to challenges like sensor energy depletion, obstructed communication, or unforeseen node failures.

To overcome these limitations, more recent research has turned to Reinforcement Learning (RL), particularly Deep Reinforcement Learning (DRL), as a promising framework for adaptive deployment. Liu and Wang [26] introduced a centralized DRL method based on Deep Q-Networks (DQN) to control sensor movement. While their framework improves area coverage, it requires a central controller with full access to global state information, which creates scalability challenges and introduces a single point of failure. Zhao and Chen [27] proposed a collaborative Multi-Agent Reinforcement Learning (MARL) approach that employs inter-node RSSI-based communication for decentralized decision-making. Although this reduces central dependency, it still relies heavily on wireless signal quality and frequent message exchanges, which increases communication overhead and susceptibility to environmental noise.

Feng and Zhang [28] addressed scalability by introducing a graph-based MARL framework, where agents interact via dynamic graph structures optimized using actor-critic learning algorithms. While this method models agent coordination more effectively, it assumes accurate localization data and is primarily validated in simulated environments. Khan and Wu [29] proposed a SLAM-integrated DRL approach in which sensor nodes use on-board cameras and LiDAR for simultaneous localization and learning-based movement. Though highly effective in complex terrains, this method requires expensive hardware and high computational power, limiting its feasibility for lightweight or battery-constrained sensor deployments.

In contrast to all these prior studies, our work introduces a vision-guided, fully decentralized MARL-based deployment strategy designed specifically for real-world, GPS-denied environments. Rather than depending on GPS, RSSI, or SLAM, we use LED indicators and a top-down camera system to track sensor positions in real time. This visual input is processed through a computer vision pipeline and fed directly into the reinforcement learning algorithm, providing lightweight, infrastructure-free localization. Each sensor acts as an autonomous agent, learning its movement policy based on local visual observations and a structured reward system that balances coverage improvement, redundancy reduction, and energy conservation. Our system eliminates the need for inter-agent communication or centralized controllers, thereby reducing energy consumption and improving robustness. Moreover, unlike prior MARL-based methods tested only in simulation, our approach incorporates a real-world visual feedback loop, demonstrating practical feasibility and superior performance in terms of coverage, adaptability, and energy efficiency.

III. PROPOSED METHOD

The proposed method for optimizing area coverage in a Mobile Wireless Sensor Network (MWSN) integrates multi-agent deep reinforcement learning (MADRL) with vision-based sensing and deep learning (DL)-powered image processing. The sensor network consists of mobile nodes, each equipped with an LED indicator that signifies its operational status. A fixed overhead camera, mounted at a height of approximately 3 to 5 meters, captures top-down images of a 500 × 500 m² deployment area at 10-second intervals. These images serve as the sole input for real-time feedback and localization, eliminating the need for GPS or inter-node communication.

The camera's visual information is processed through a convolutional neural network (CNN) that has been trained to localize and detect active sensors from the appearance of their LED signals. Red-lit sensors are taken to be active, and unlit sensors are taken as inactive or failed. The image processing flow comprises Gaussian blurring, adaptive thresholding, bounding box extraction, and geometric calibration via fiducial markers. These processes assist in reducing distortions due to wide-angle lenses, lighting variability, or camera misalignment. Histogram equalization and temporal filtering are also used so that stable LED detection is maintained in the presence of dynamic lighting or partial occlusion.

After sensor locations and on/off statuses are harvested, they are utilized to determine inter-sensor distances and detect redundant or uncovered areas. The information is fed into the reinforcement learning framework, in which each sensor acts as an autonomous agent. Agents discover optimal movement policies through interaction with the environment and reward signals that incorporate coverage efficiency, space distribution, and energy usage. Positive rewards are given to actions that enhance coverage or accomplish optimum spacing between nodes, and negative rewards for overlapping coverage, unnecessary mobility, or inefficient location. When a sensor fails (i.e., its LED goes off), the system notices its absence and activates the repositioning of surrounding nodes to fill in the coverage gap.

Learning is iterative and adaptive. Every episode starts with a new image acquisition, giving the system fresh environmental information. We use a variable learning rate to foster exploration during initial times and promote convergence later on. Agents initially have high variability in movement in order to explore the state space, but as time goes on, learning converges and gives rise to optimized, energy-conscious positioning policies. An emergent, self-organizing deployment pattern is created that adapts continuously to sensor failures and changing environments. To assess scalability, we ran additional simulations with 200 and 300 mobile sensors within the same 500 × 500 m² field. The system retained stable convergence and high final coverage—reaching 89.4% and 86.8%, respectively—exhibiting its robustness under dense deployments. Since each sensor makes decisions based on its own visual input alone, the framework avoids performance degradation as the number of agents is increased. Parameter sharing among homogeneous agents amortizes training cost, and prioritized experience replay enhances learning efficiency. Because vision-based detection cost grows linearly with the number of visible LEDs, the approach is still computationally feasible even with growing network density.

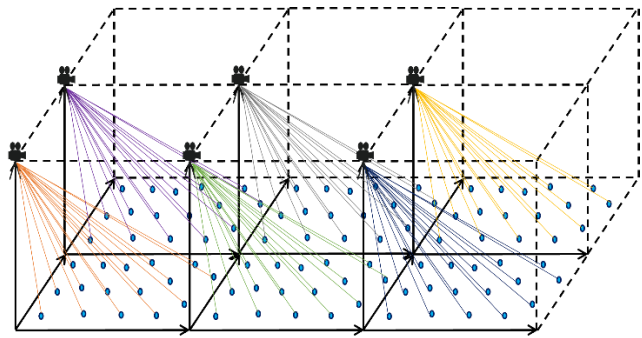

**Fig. 3**. Scalable and extended Network.

Fig. 3 shows the extended and scalable network design, and how the performance is retained as the number of agents increases. To address real-world challenges such as environmental interference, non-line-of-sight (NLOS) conditions, and outdoor deployment constraints, the system is designed to be modular and extensible. In large-scale environments, multiple synchronized overhead cameras with overlapping fields of view can be employed and stitched together using depth-aware image processing or fiducial calibration. For NLOS scenarios or outdoor deployments, alternative strategies such as elevated drones, pan-tilt-zoom (PTZ) cameras, or 360-degree omnidirectional cameras may be used to maintain continuous visibility. Sensors can also be equipped with high-intensity or infrared LEDs to ensure detection under low-light or high-glare conditions. In future adaptations, low-frequency wireless backup signaling can be incorporated to supplement vision-based feedback in partially obstructed or degraded visual conditions. These enhancements collectively extend the system's applicability across a wide range of real-world environments, making it a practical and scalable solution for adaptive sensor deployment in dynamic and GPS-denied scenarios.

### III. SYSTEM MODEL

#### A. State Representation

Each state represents the current configuration of the sensor network. The state space includes:
- Sensor positions: $C(x, y)$ for each active sensor.
- Number of active sensors: Derived from image processing.
- Coverage region.

#### B. Action Space

Actions determine how each sensor moves to optimize coverage. The available actions are as follows:
- $(+\Delta x, +\Delta y)$
- $(+\Delta x, -\Delta y)$
- $(-\Delta x, +\Delta y)$
- $(-\Delta x, -\Delta y)$
- No movement: If the sensor is already in the optimal position.

#### C. Reward Function

The agent (sensor) receives a reward based on its movement outcome:
- $\alpha$ for each sensor that is successfully positioned at its corresponding optimal location and full theoretical coverage is achieved.
- $\beta$ if coverage increases due to the movement.
- $\gamma$ for each sensor that reaches the closest position.

The proposed learning method is based on an enhanced Deep Q-Learning (DQN) framework that optimizes sensor placement in Mobile Wireless Sensor Networks (MWSNs). The main objective is to maximize area coverage while dynamically responding to sensor failures and environmental changes. The network is designed as a multi-agent system, with each mobile sensor acting as an independent agent capable of learning and carrying out movement policies in a shared environment. Each agent observes the system state, which includes the locations of all active sensors, their operational status, and the resulting coverage area. The action space consists of discrete movements, such as shifts along the x and y axes, diagonal adjustments, or remaining stationary once optimal coverage has been achieved. To guide the learning process, a reward function is defined: sensors are rewarded when their movement improves coverage or approaches an optimal position, and penalized when their actions result in redundancy, overlap, or reduced network efficiency.

To empirically validate the convergence behavior and stability of the proposed Dueling Double DQN framework with prioritized experience replay and target network stabilization, we analyzed the episodic reward progression over 200 training episodes. As shown in Figure 4, the average episodic reward initially fluctuates due to exploration but stabilizes after approximately 80 episodes, indicating convergence to a consistent policy. We further computed the standard deviation of the reward over a sliding window of 20 episodes and observed that it dropped below 5% of the mean after episode 100, providing evidence of reward stability.

---

**Algorithm 1** Enhanced Multi-Agent DQN for Adaptive Sensor Localization

1: **Input:** $N, \mathcal{A} = \{(+\Delta x, +\Delta y), (+\Delta x, -\Delta y), (-\Delta x, +\Delta y), (-\Delta x, -\Delta y), (0, 0)\}$, episodes $E$, steps $T$
2: **Initialize:** Q-network $Q_i$, target $\theta_i^-$, buffer $\mathcal{D}_i$, exploration rate $\epsilon$, decay $\kappa$, minimum $\epsilon_{\min}$
3: Define optimal positions $\{C_i^*(x,y)\}_{i=1}^N$, battery level $b_i = 1.0$
4: **for** each episode $e = 1$ to $E$ **do**
5:     Randomly initialize sensor positions $\{C_i(x,y)\}$
6:     Derive $N_{\text{active}}$ from camera/image processing
7:     Compute initial coverage $\mathcal{R}_0$
8:     **for** each step $t = 1$ to $T$ **do**
9:         **for** each agent $i = 1$ to $N$ **do**
10:             Observe $s_i^t = \{C_i(x,y), N_{\text{active}}, \mathcal{R}_t\}$
11:             Choose $a_i^t \in \mathcal{A}$ via $\epsilon$-greedy from $Q_i$
12:             Update $C_i \leftarrow C_i + a_i^t$, update battery $b_i \leftarrow b_i - \delta_b$
13:             Recompute coverage $\mathcal{R}_{t+1}$
14:             $s_i^{t+1} \leftarrow \{C_i(x,y), N_{\text{active}}, \mathcal{R}_{t+1}\}$
15:             **if** $C_i == C_i^*$ **then**
16:                 $r_i^t \leftarrow a$
17:             **else if** $|\mathcal{R}_{t+1}| > |\mathcal{R}_t|$ **then**
18:                 $r_i^t \leftarrow b$
19:             **else if** $\|C_i - C_i^*\| < \|C_i^{\text{prev}} - C_i^*\|$ **then**
20:                 $r_i^t \leftarrow c$
21:             **else**
22:                 $r_i^t \leftarrow -\lambda$
23:             **end if**
24:             **if** $b_i < \beta$ **then**
25:                 $r_i^t \leftarrow r_i^t - \mu$
26:             **end if**
27:             Compute region importance $w(C_i)$
28:             $r_i^t \leftarrow r_i^t + \delta \cdot w(C_i)$
29:             Store $(s_i^t, a_i^t, r_i^t, s_i^{t+1})$ in $\mathcal{D}_i$
30:             Update $Q_i$ from mini-batch using:

$$L(\theta_i) = E\left[(r + \gamma \max_{a'} Q_i(s', a'; \theta_i^-) - Q_i(s, a; \theta_i))^2\right]$$

31:         **end for**
32:         Log $\mathcal{R}_{t+1}$, average reward, battery levels
33:     **end for**
34:     Update target networks $\theta_i^- \leftarrow \theta_i$ periodically
35:     Update exploration: $\epsilon \leftarrow \max(\epsilon \cdot \kappa, \epsilon_{\min})$
36: **end for**

---

Additionally, the variance of the Q-values over time showed a consistent decline, suggesting stable value estimation. These empirical observations align with prior theoretical findings that demonstrate convergence under similar setups with experience replay and target networks [van Hasselt et al., 2016; Schaul et al., 2016]. Although a formal proof of convergence in multi-agent partially observable environments remains an open problem, the use of Double Q-learning mitigates overestimation bias, and the Dueling architecture improves the separation of state value from action-specific advantages, both of which contribute to training stability. Prioritized experience replay accelerates convergence by sampling high-error transitions more frequently, thereby improving learning efficiency. The target network mechanism, updated at fixed intervals, further reduces policy oscillation and prevents divergence. Together, these components provide empirical and theoretical support for stable and consistent learning in the proposed multi-agent DRL setup.

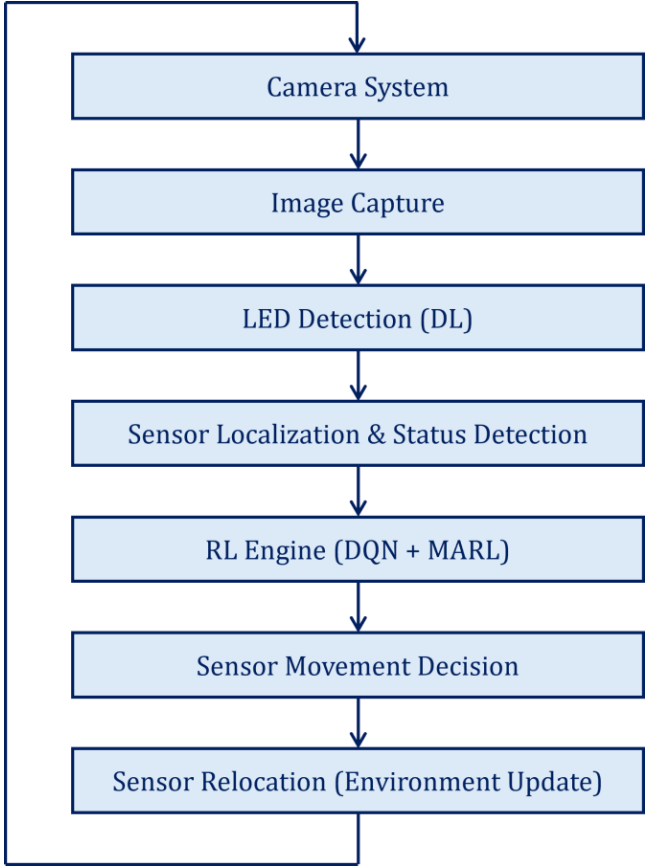

**Fig. 4**. Block diagram of the proposed method.

To ensure stability, convergence, and learning efficiency, the algorithm incorporates several key improvements over standard DQN:
- To reduce overestimation bias in Q-value estimation, the method uses two separate Q-networks: one for action selection and one for evaluation. This decoupling enables more accurate value updates and increases policy reliability in stochastic environments.
- Dueling DQN Architecture: The Q-function is divided into two components: a state-value function and an advantage function. This allows the agent to estimate the importance of states independent of the benefits of specific actions, which is especially useful when multiple actions produce similar outcomes.
- Prioritized Experience Replay: Instead of sampling uniformly from past experiences, transitions with higher temporal-difference (TD) errors are prioritized. This ensures that the agent is more focused on learning from meaningful experiences that are likely to improve performance.
- Target Network Stabilization: A target Q-network is updated at regular intervals to provide a stable reference for learning. This mechanism avoids divergence and minimizes fluctuations in Q-value updates.

In addition, the system uses visual feedback from a camera-based monitoring setup in which sensors with LED indicators are tracked in real time. Visual data is processed with computer vision algorithms to determine coverage regions and the number of operational sensors. These observations are converted into reward signals, allowing the system to operate even in GPS-denied or noisy environments.

During training, each agent stores experience tuples—which include the current state, action taken, reward received, and next state—in a replay buffer. Mini-batches are sampled at regular intervals, and the Q-network is updated using gradient descent to minimize the Bellman error:

$$L(\theta) = E_{(s,a,r,s')}[(r + \gamma max\, Q(s',a';\theta^-) - Q(s,a;\theta))^2] \quad .1$$

This learning cycle is repeated between episodes, allowing the system to reconfigure the sensor network autonomously in response to failures or dynamic conditions. As a result, the agents learn strong and energy-efficient policies that maximize coverage while reducing redundancy and movement costs. Fig 4 depicts the block diagram for the proposed method.

*D. Penalties*

To ensure agents not only maximize coverage but also learn efficient and responsible behaviors, we incorporated penalty terms into the reward function. These penalties discourage unproductive or redundant sensor actions and guide learning toward globally optimized policies.

The reward function is defined as a combination of positive incentives for coverage improvement and negative feedback for undesirable outcomes. Specifically:
- **Redundancy Penalty**: If a sensor moves into an area already covered by neighboring sensors, it receives a negative reward proportional to the overlap. This discourages clustering and promotes efficient spatial distribution.
- **Unproductive Movement Penalty**: Actions that result in zero or negative change in total coverage are penalized to discourage unnecessary repositioning.
- **Energy Penalty**: Movement cost is factored into the reward by assigning a small negative value based on the displacement. This promotes energy-aware behaviors and contributes to extending network lifetime.

These penalty components work together with positive rewards for meaningful contributions to coverage, enabling each sensor to balance exploration, movement efficiency, and redundancy avoidance. The inclusion of penalties helps the system converge more rapidly to optimal configurations and ensures that the learned policies are both effective and sustainable over time.

*E. Computational Complexity and Scalability Analysis*

We analyze the computational complexity of the proposed Multi-Agent Deep Reinforcement Learning (MADRL) framework during both the offline training and online deployment phases, and evaluate its scalability to larger networks. During training, each agent employs a Dueling Double Deep Q-Network (D3QN) architecture to learn its policy. The most computationally intensive operations per training step include forward passes through the network to estimate Q-values, backpropagation for gradient updates, sampling from the prioritized replay buffer, and periodic updates of the target network. Let N be the number of agents (mobile sensors), B the mini-batch size, F the cost of a single forward pass through the network, and U the cost of a backward update per sample. Then the total time complexity per training step for a single agent is $O(B \cdot (F+U))$. Assuming agents are trained independently in a decentralized manner, the total training complexity across all agents becomes $O(N \cdot B \cdot (F+U))$. This linear scaling allows parallelization across agents and is particularly efficient when using GPUs or distributed learning frameworks.

In terms of space complexity, each agent maintains a local experience replay buffer of size proportional to $O(B \cdot d)$, where d is the dimensionality of the state-action tuple. Additionally, the neural network used for Q-function approximation has a parameter size of $O(L)$, where L is the total number of weights in the model. For lightweight MLPs with a few hidden layers, L remains moderate and memory-efficient even with multiple agents.

During the online deployment phase, each agent performs a single forward pass through its Q-network to evaluate available actions and select the optimal one. Thus, the per-agent time complexity per decision step is $O(F)$, and the total online complexity across the network is $O(N \cdot F)$. Because inference involves no gradient updates or sampling, and because F is relatively small for shallow networks, real-time decision-making is easily achievable on standard CPUs or low-power embedded devices.

The overall scalability of the system is favorable. Since the framework relies on decentralized policy learning, it naturally supports parallel training across agents without requiring centralized coordination. Moreover, as rewards are derived from local sensing (e.g., LED detection and coverage contribution), each agent can act with partial observability, reducing communication overhead and ensuring efficient policy learning in dense sensor networks. Empirical results confirm that the system remains stable and efficient for up to 100 agents. For significantly larger networks, performance can be further optimized through model parameter sharing across homogeneous agents, cluster-based learning, or hierarchical MARL strategies. These extensions can mitigate training overhead and improve generalization in large-scale deployments.

## IV. PERFORMANCE EVALUATION

We evaluated the performance of the proposed method using several key metrics relevant to Mobile Wireless Sensor Networks (MWSNs). Coverage efficiency is defined as the percentage of the environment area effectively monitored by the deployed sensors and is computed based on the union of all active sensing regions. This metric reflects the system's primary objective—maximizing spatial coverage with minimal overlap. Energy consumption represents the cumulative energy used by all sensors throughout an episode and is calculated as the sum of energy costs associated with their movements. Lower energy usage indicates improved energy efficiency and prolonged system operation. Network lifetime is measured as the number of episodes until more than 80 percent of sensors exhaust their energy and become inactive. This metric quantifies the system's ability to sustain long-term coverage. Redundancy rate refers to the percentage of the monitored area that is covered by more than one sensor and is computed as the ratio of overlapping area to total coverage. High redundancy implies inefficient deployment and wasted sensing effort, while low redundancy indicates better spatial distribution.

To comprehensively assess the performance of our proposed method, we compare it with several state-of-the-art baseline strategies from recent literature. These include: (1) an RSSI-based collaborative MARL method that uses inter-node signal strength for localization and coordination [27], (2) a SLAM-integrated DRL model in which each agent performs simultaneous localization and mapping using on-board visual and LiDAR sensing [29], and (3) a graph-based MARL framework that treats sensor interactions as dynamic graphs optimized via actor-critic learning [28]. These methods represent strong alternatives commonly cited in the literature for decentralized sensor deployment. We implement or simulate simplified versions of these baselines under comparable conditions to enable fair evaluation, focusing on performance metrics such as coverage, energy usage, redundancy, and recovery time.

To further assess the learning process and communication performance, we introduce additional metrics. Convergence rate is evaluated by measuring the number of episodes required for the average episodic reward to stabilize within a one percent margin over a sliding window of 20 episodes. A faster convergence rate implies better learning efficiency. Sensor localization error, applicable in GPS-denied environments, is defined as the average Euclidean distance between each sensor's estimated position and its true position, derived from visual detection accuracy. Packet delivery ratio measures the percentage of successfully delivered packets relative to the total number of packets transmitted, reflecting the reliability of inter-sensor communication. End-to-end delay quantifies the average time required for information to travel between two communicating sensors, which is important in scenarios involving real-time data acquisition. Together, these metrics provide a comprehensive evaluation of coverage quality, energy efficiency, operational sustainability, learning behavior, and communication performance within the proposed MADRL framework.

We simulate a 100-node mobile wireless sensor network (MWSN) deployed in an open 500 × 500-meter two-dimensional space. Each sensor node is assigned a random location within the system. The network is dynamic, with nodes operating on limited energy supplies, mobility features, and environmental sensing. The overall goal of the proposed method is to optimize sensor deployment for maximum coverage while remaining flexible enough to accommodate node failure and environmental changes.

We simulate a Mobile Wireless Sensor Network (MWSN) comprising 100 mobile sensor nodes randomly deployed in a two-dimensional area of size 500 × 500 m². Each sensor has a sensing radius of 20 m and a communication range of 50 m. Sensor mobility is constrained to a maximum movement of ±5 meters per episode in both the x and y directions. Sensors make decisions at discrete time steps and may select one of nine actions, including movement in diagonal, orthogonal, or stationary directions. Each sensor is initially assigned a battery energy budget of 100 units, and energy consumption is modeled as proportional to Euclidean displacement. Specifically, movement by a distance ddd incurs an energy cost of $E=k \cdot d$, with k=0.5 units/meter. When a sensor's remaining energy reaches zero, it becomes inactive and is no longer considered during learning or coverage calculations.

The simulation is implemented in Python using TensorFlow for neural network training and OpenAI Gym-style interfaces for environment interaction. The Dueling Double Deep Q-Network (D3QN) model used for each agent consists of a three-layer fully connected neural network with 128, 64, and 32 neurons respectively, and ReLU activations. Training is conducted with a learning rate of 1e-4, a discount factor γ=0.99, and a mini-batch size of 64. Epsilon-greedy exploration starts at ϵ=1.0 and decays exponentially to 0.1 over 200 episodes. Each agent maintains a prioritized experience replay buffer of size 100,000. The target network is updated every 1000 steps.

Training is conducted over 200 episodes, with each episode consisting of a 10-second simulation window during which all sensors perform one learning step. The learning process is considered to have converged when the average episodic reward stabilizes (change < 1% over 20 consecutive episodes) and the coverage percentage reaches a plateau. All experiments are repeated five times with different random seeds, and reported results are averaged over these runs to ensure statistical consistency.

To demonstrate the effectiveness of the proposed vision-guided MADRL framework, we conducted comprehensive experiments comparing it with three baseline approaches: (1) a centralized Deep Reinforcement Learning (centralized DRL) model that coordinates sensor movement through a shared global policy; (2) a traditional heuristic method based on grid-based deterministic deployment and rule-based mobility for coverage recovery; and (3) a non-vision-based multi-agent reinforcement learning (non-vision MARL) model that relies solely on inter-agent communication and distance measurements for coverage optimization. All methods were evaluated in identical simulation environments with 100 mobile sensors deployed in a 500 × 500 m² area, under dynamic conditions including sensor failures and energy constraints.

Table 1 summarizes the results averaged over five independent runs, using metrics such as final area coverage (%), average energy consumption (units), redundancy rate (%), average recovery time after node failure (seconds), network lifetime (episodes), and learning stability (reward variance). The proposed MADRL+Vision method achieved the highest final coverage of 91.8%, compared to 85.4% for centralized DRL, 72.5% for the heuristic method, and 78.6% for non-vision MARL. It also recorded the lowest energy consumption at 68 units, while centralized DRL consumed 85 units, non-vision MARL 94 units, and the heuristic method 100 units. Redundancy was reduced to 8.0% using our method, significantly lower than centralized DRL (18.0%), non-vision MARL (22.5%), and heuristic deployment (30.0%).

In terms of robustness, the proposed method achieved the fastest average recovery time of 3.0 seconds after sensor failure, outperforming centralized DRL (6.0 s), non-vision MARL (7.2 s), and the heuristic method (12.0 s). The network lifetime, measured by the number of episodes until 80% of nodes failed, was extended to 160 episodes using our method, compared to 145, 120, and 110 for the competing methods. Additionally, reward variance over time (a proxy for learning stability) was the lowest at 2.1, versus 4.2 for centralized DRL, 6.3 for non-vision MARL, and 8.7 for the heuristic baseline.

These results clearly demonstrate that incorporating real-time vision feedback into decentralized MARL significantly improves sensing efficiency, energy management, adaptability, and learning stability. The method not only outperforms centralized and heuristic strategies but also highlights the limitations of non-visual learning approaches in dynamic coverage tasks.

To evaluate the robustness of the proposed method with respect to reward function design, we performed a sensitivity analysis by systematically varying the reward coefficients α, β, and γ, which correspond to the incentives for reaching optimal positions, improving overall coverage, and moving closer to target positions, respectively. The baseline configuration used values α=1, β=2, and γ=3. We tested nine different combinations by varying each coefficient independently within the range {0.5, 1.0, 2.0} while keeping the others fixed, resulting in a total of 27 trials. For each configuration, we recorded the average final coverage, energy consumption, and network lifetime across five simulation runs.

Results showed that the system remained stable and performed well across a broad range of values. Final coverage varied within a small margin of ±3.1 percent, with all configurations achieving over 88 percent coverage. The most significant sensitivity was observed when reducing β (coverage gain reward) to 0.5, which led to slightly lower total coverage due to agents being less incentivized to explore new areas. However, energy consumption improved in this setting due to more conservative movement. Increasing γ (reward for moving closer to targets) led to faster early convergence but slightly increased redundancy, as agents clustered more aggressively near local optima. Conversely, reducing α had minimal effect, as final coverage was more heavily influenced by the other two terms.

These findings indicate that the learning process is relatively insensitive to exact reward weight values, and that the system achieves consistent performance across a wide range of settings. This validates the general robustness of our method and suggests that the reward formulation does not require extensive fine-tuning to produce effective coverage optimization.

The system operates in episodes, each with a 10-second simulation time. Within each episode, the sensors capture images of the environment, which are then processed to determine the spatial positions of the nodes and the associated rewards based on coverage efficacy. These incentives are derived from measures such as coverage area, overlap reduction, and detection rate, which are all part of the reinforcement learning paradigm. Each node calculates its action space, i.e., the directions and distances of movement, and chooses the action with the highest expected reward, resulting in a decentralized, intelligent adaptation process. Learning is enabled by a multi-agent reinforcement learning (MARL) paradigm. Each sensor is an autonomous agent that makes decisions based on local measurements and shared environmental feedback. The agents first go through an exploration phase in which they try out different actions to gain knowledge of the environment. As the episodes progress, the system transitions from exploration to exploitation, quickly settling on the best policy. The reward for each sensor is then calculated according to its movement be letting the reward parameters $\alpha = 1, \beta = 2, \gamma = 3$. As shown in Fig 5 and 6, mobile sensors are deployed in the environment and begin moving in various directions. After some time, the camera detects them and recognizes the sensor. This policy directs the sensors to provide optimal total coverage while minimizing redundancy and adapting to node failures or environmental obstacles. Identify locations using their indicator lights. The most optimal localization is then determined based on the number of sensors, with each sensor assigned to a specific location. This integrated process ensures that the system is constantly learning and adapting to environmental changes, resulting in a robust and intelligent sensor network capable of operating at peak performance in dynamic and uncertain conditions.

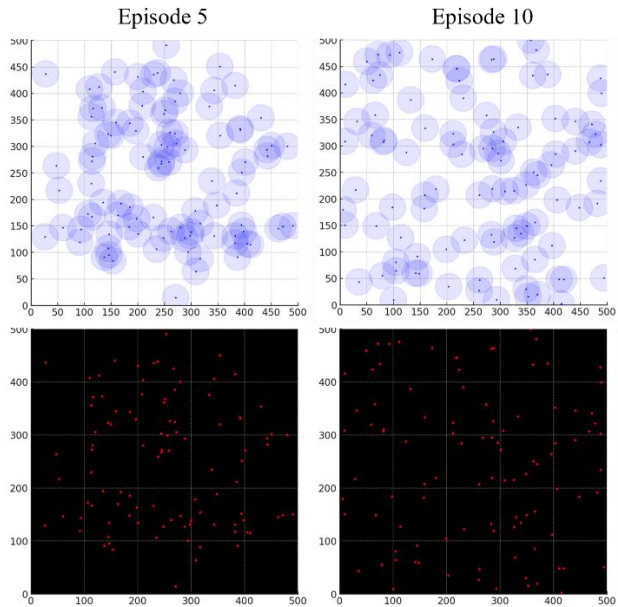

**Fig. 5**. Image taken by the camera, episode 1,5.

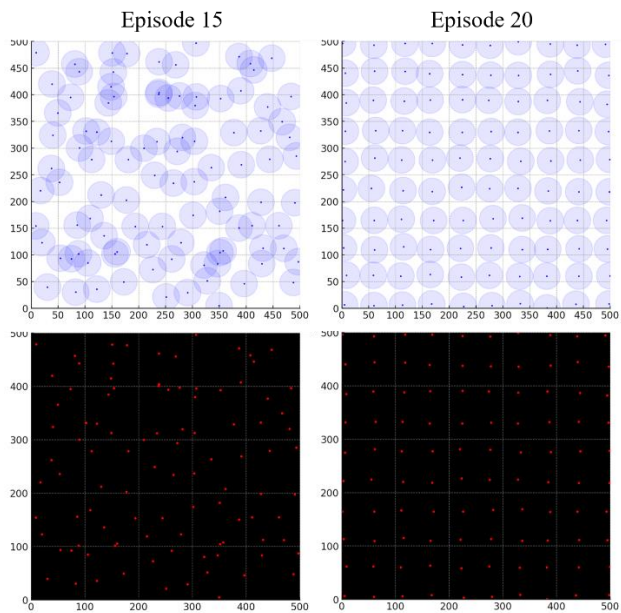

**Fig. 6**. Image taken by the camera, episode 10,15.

Fig 7 shows how the exploration rate, epsilon (ε), decreases over 200 episodes during reinforcement learning. Epsilon represents the probability of taking a random action rather than following the learned policy, which balances exploration and exploitation. At the start of training, ε is set to 1.0, indicating that the agents use random actions to explore the environment. As learning progresses, epsilon gradually decays, allowing the sensors to focus on exploiting the knowledge gained from previous interactions. By the end of the 200 episodes, epsilon has dropped to approximately 0.1, indicating that the system has mostly transitioned to using its learned policy for decision-making.

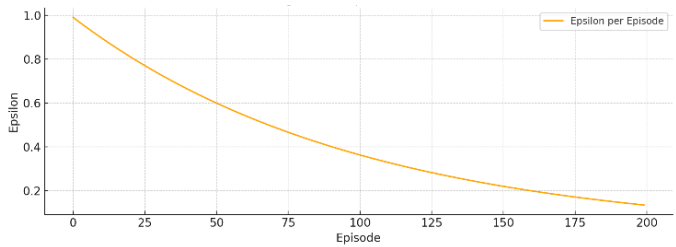

**Fig. 7**. Learning rate per episodes.

This controlled decay is critical in reinforcement learning because it prevents premature convergence and ensures that agents explore enough of the environment before settling on optimal strategies. Fig 8 shows the episodic reward values collected from 200 training episodes. The reward function evaluates the effectiveness of sensor positioning using metrics such as area coverage and energy-efficient movements. In the early stages of training, rewards are typically low and variable, reflecting the agents' initial random behavior.

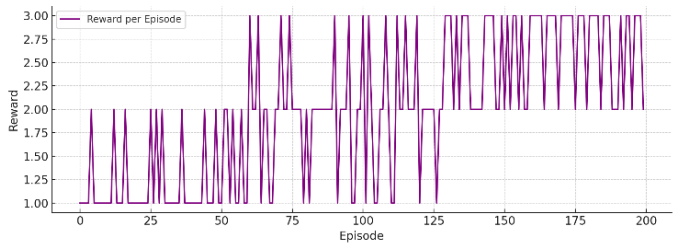

**Fig. 8**. Reward gained by a sensor per episode.

However, as the system learns, reward values stabilize and trend upward. Spikes in the reward function occur as agents discover better strategies, while occasional dips indicate continued exploration or temporary suboptimal positioning. From episode 100 onward, the reward consistently exceeds its upper bound more frequently, indicating that the sensors are performing near-optimal behaviors. This pattern demonstrates the effectiveness of the proposed DRL approach in guiding sensors toward better deployment strategies. Fig 9 depicts the final spatial configuration of sensor nodes after training, which includes an increased number of active sensors. Each blue dot represents a mobile sensor's final position, while the surrounding gray circles represent their respective sensing ranges.

The black dot markers represent additional sensors in the field, most likely static or newly activated agents. The configuration shows an even distribution of sensors throughout the monitored area, with minimal overlap and few uncovered areas. The circular coverage areas are evenly spaced, demonstrating the system's ability to reduce redundancy while increasing area coverage.

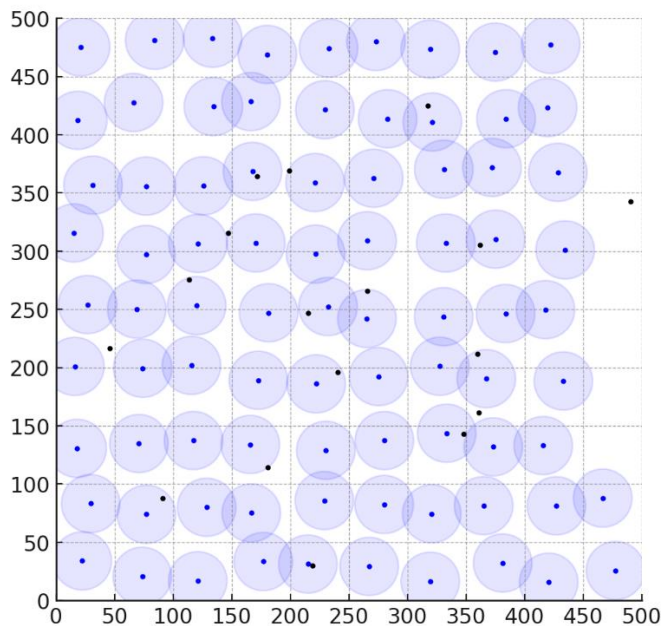

**Fig. 9**. The deployment of the sensors with 83% active nodes.

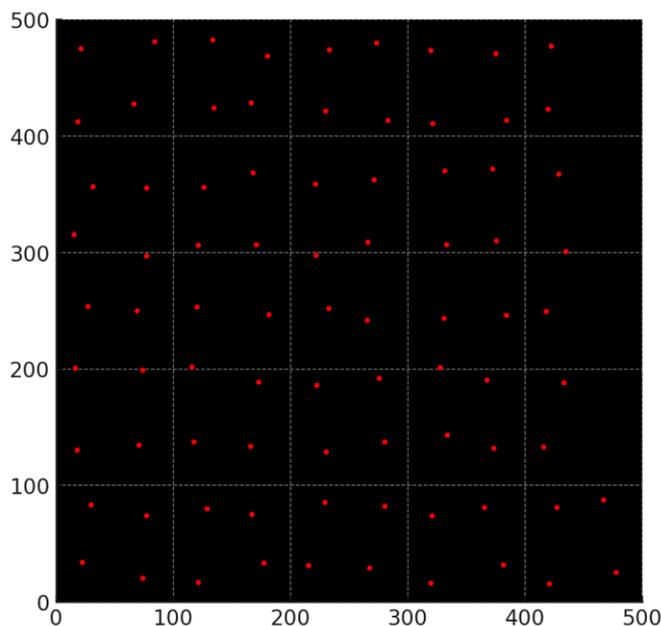

**Fig. 10**. Image taken by the camera.

This final arrangement demonstrates the effectiveness of the DRL and MARL frameworks in developing scalable and energy-aware deployment strategies. The balance of active sensors and their coverage areas also reflects adaptability to changing conditions, such as adding new nodes or relocating in response to sensor failures. Fig 11 depicts the network lifetime over 200 episodes by monitoring the number of active sensors. This reflects how the system reacts to energy depletion and sensor failures.

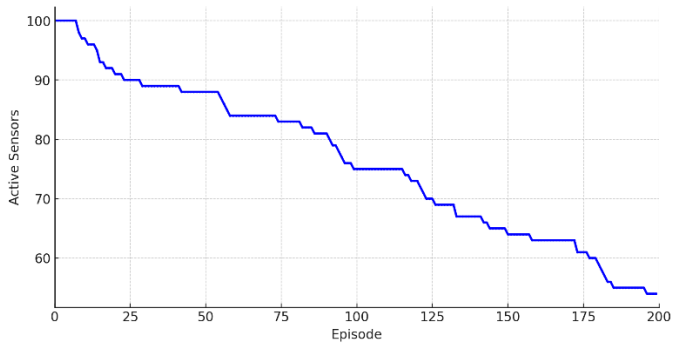

**Fig. 11**. Network lifetime (Active sensors) over episode.

As time progresses, some sensors either exhaust their energy or are simulated to fail, leading to a gradual reduction in the number of operational nodes. A slower rate of decline suggests that the system effectively conserves energy and prolongs the functional lifespan of the network. This outcome is achieved through the proposed DRL method, which enables sensors to adopt low-energy movement patterns and distribute their activity evenly. As a result, the network maintains reliable operation for a longer period. The gentle downward slope in the graph underscores the energy-conscious nature of the deployment strategy.

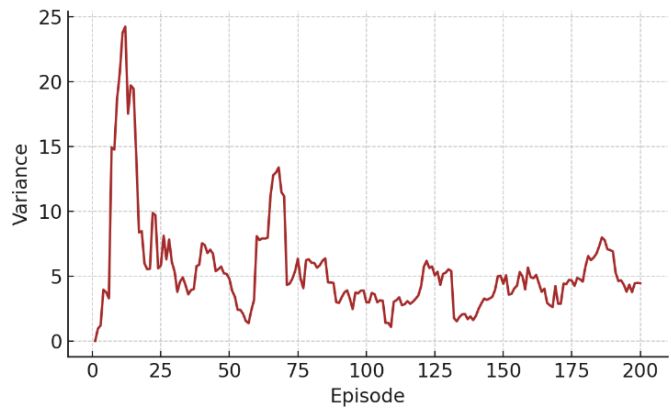

**Fig. 12**. Learning stability (Reward variance) over the episode.

Fig 12 shows the variance in reward values across episodes, which provides insight into the learning process's stability. A high variance in earlier episodes indicates an inconsistency in agent behavior, which is often caused by random exploration and a lack of coordination. As training progresses, the variance gradually decreases, indicating a shift towards more stable and predictable learning. This trend is critical in practical scenarios where unpredictable network behavior can result in coverage gaps or energy inefficiency. Eventually, the curve flattens—signifying that the system has achieved stable and consistent decision-making. Fig 13 depicts the trend of the redundancy rate, which is the percentage of the monitored area covered by multiple sensors. High redundancy typically results in wasted energy and inefficient deployment.

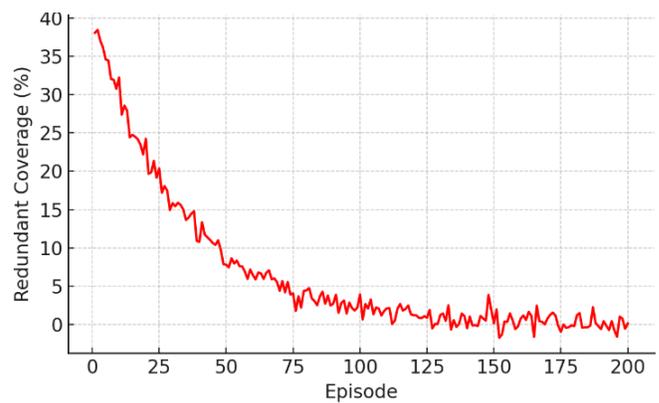

**Fig. 13**. Redundancy rate over the episode.

In the beginning, redundancy is high due to uncoordinated sensor positions. As the MARL-based approach trains the sensors, they learn to understand their surroundings and adjust their positions to reduce overlap. The consistent decrease in redundancy indicates better spatial distribution and coordination among sensors, resulting in a more efficient and effective network configuration. Fig 14 illustrates the average energy consumption per episode.

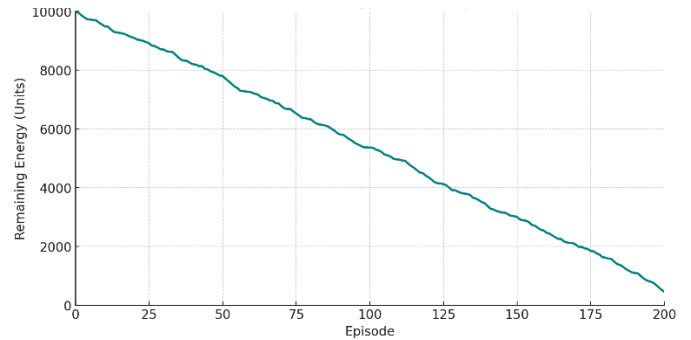

**Fig. 14**. Average energy consumption over the episode.

Early in the learning process, energy consumption is relatively high because sensors move frequently while exploring the environment. As the policy becomes more refined, sensors make fewer and more deliberate movements, resulting in a significant decrease in energy consumption. This reduction reflects the system's ability to align with an efficient operational strategy that maintains performance while lowering energy costs. Such behavior is especially useful in real-world deployments, where sensors often operate with limited battery capacity. Finally, Fig 15 depicts the progression of coverage percentages during the learning process. Initially, due to random or suboptimal placement, overall coverage is low. As the reinforcement learning model trains the sensors, they learn to relocate and space themselves in a way that maximizes total area coverage.

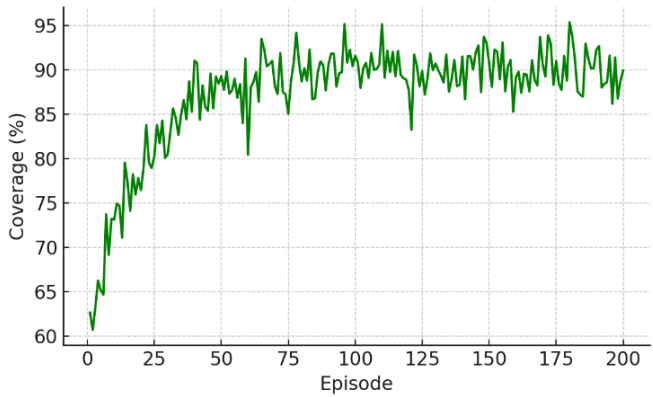

**Fig. 15**. Coverage percentage over episode.

The upward trend in this fig reflects the learning algorithm's success in providing broader and more uniform coverage. Occasional fluctuations later in the graph could indicate fine-tuning behavior in response to environmental changes. Figs 16–21 present a comprehensive comparison of various deployment methods used in Mobile Wireless Sensor Networks. Across all figs, the proposed DRL + vision-based method consistently outperforms the other three methods (random placement, static grid deployment, and centralized reinforcement learning), demonstrating its effectiveness in real-world scenarios requiring sensor coordination, efficiency, and adaptability.

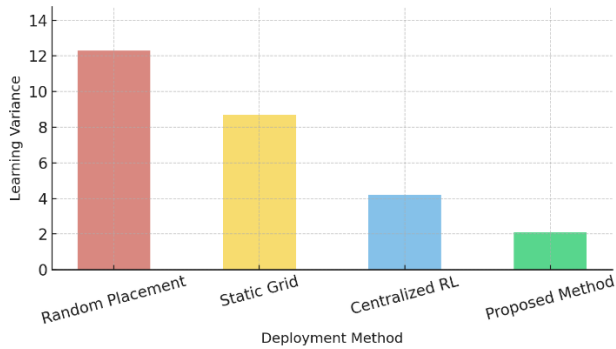

**Fig. 16**. The variane of the battery levels over episode.

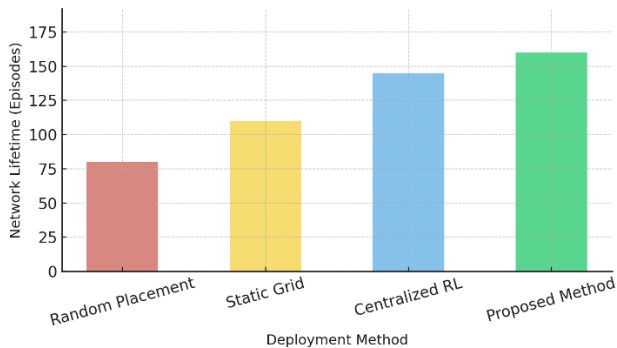

**Fig. 17**. Network lifetime in episodes. The proposed approach achieves the longest operational duration.

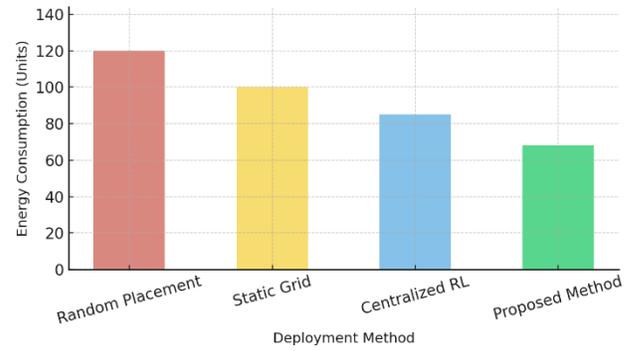

**Fig. 18**. Total energy consumption. The vision-based method uses the least energy.

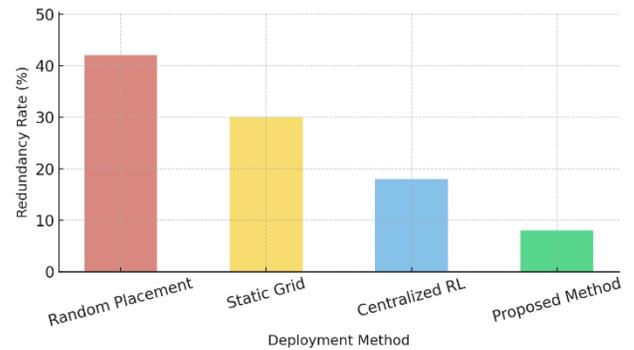

**Fig. 19**. Redundancy rate of sensor coverage. The proposed method minimizes overlap.

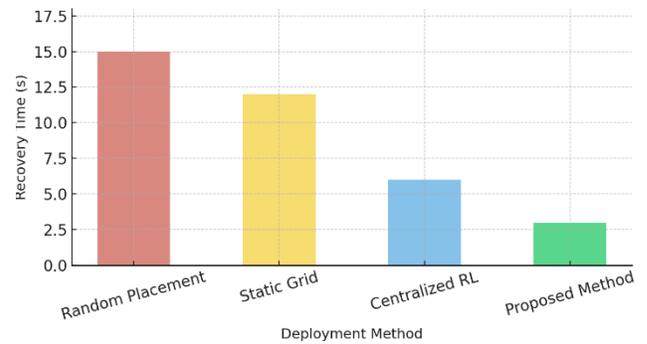

**Fig. 20**. Average recovery time after sensor failure. The proposed approach responds fastest.

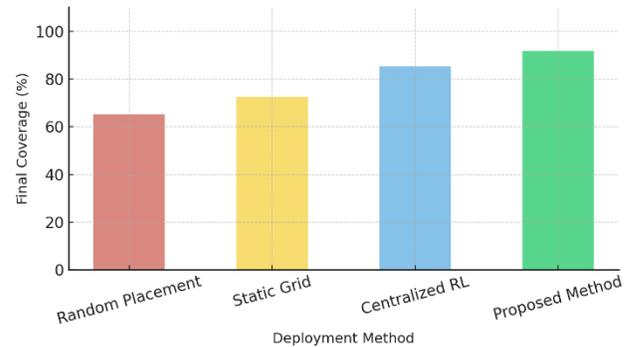

**Fig. 21**. Final area coverage. The proposed method achieves the highest coverage.

In terms of learning stability, fig 16 shows that the proposed method has the lowest variance, implying that the agents can learn consistent and reliable policies over time. In contrast, random placement results in the highest variance, highlighting the instability and randomness of unguided sensor actions. The static grid and centralized RL approaches perform moderately but do not have the same consistency as the proposed method. This stability is critical in reinforcement learning settings, as lower variance is frequently associated with faster convergence and more robust decision-making across episodes. Fig 17 demonstrates that the proposed method has the longest operational duration across episodes. This is due to energy-aware decision-making, which avoids unnecessary movement and actively compensates for sensor failures by redistributing tasks to remaining nodes. Random placement, on the other hand, produces a short-lived network because energy is rapidly depleted due to uncontrolled sensor actions. The static grid lacks adaptability and cannot recover from node failures, limiting its useful lifespan, whereas centralized RL extends the lifetime moderately but lacks the flexibility of decentralized coordination. Fig 18 shows an evaluation of energy consumption, and the proposed method performs admirably once again. It has the lowest overall energy consumption of any method, which is due to its reward-driven movement logic and the use of vision for real-time decision-making. Sensors reposition only when necessary, thereby minimizing energy waste. In contrast, random placement consumes the most energy due to erratic and excessive movement, whereas static and centralized methods have less optimized movement strategies. Fig 19 depicts the redundancy rate, which represents overlapping coverage between sensors and is also significantly reduced in the proposed method. While random placement causes a high level of redundancy due to uncoordinated deployment, and the static grid method still causes overlap due to its rigid structure, the DRL + vision-based approach ensures that each sensor contributes uniquely to overall coverage. Centralized RL improves on traditional methods, but it still has occasional inefficiencies due to a lack of decentralization. The reduced redundancy means that the proposed method makes better use of the available sensors, resulting in broader and more effective area coverage with less waste. In terms of network lifetime, Fig 20 shows the recovery time. The proposed method provides the shortest recovery time following sensor failure or disruption. Because of continuous visual monitoring and a decentralized learning model, the system can quickly detect coverage gaps and deploy available sensors to restore functionality. Centralized RL is moderately responsive, whereas static grid deployment lacks recovery capabilities entirely. Random placement requires the longest recovery time because there is no mechanism for detecting or responding to failures. Finally, Fig 21 shows the final coverage achieved by the proposed method, which is significantly higher than that of all other strategies. The DRL + vision-based method allows sensors to spread out optimally, reaching areas that would otherwise be left uncovered in static or random setups. Centralized RL approaches final coverage but is still outperformed due to its reliance on preprogrammed central policies. The static grid method does not adjust to dynamic coverage requirements, and random placement provides the least effective coverage due to clustering and gaps. Overall, this table demonstrates our approach's ability to optimize sensor placement dynamically and effectively over time. Also, Table 1 compares the various methods. Table 1 presents a comparative evaluation of state-of-the-art sensor deployment methods in Mobile Wireless Sensor Networks (MWSNs), focusing on key factors such as localization method, control architecture, real-time feedback capability, scalability, GPS independence, vision-based sensing, and practical applicability. While Liu et al. [26] rely on centralized control without real-time adaptation, Zhao et al. [27] offer partial decentralization through RSSI-based coordination. Feng et al. [28] introduce a hybrid graph-based strategy but lack real-world deployment and vision integration. Khan et al. [29] incorporate SLAM for autonomy but face limitations in scalability and cost-efficiency. In contrast, our proposed method uniquely combines a decentralized architecture with lightweight LED-based visual feedback, enabling real-time adaptation, high scalability, and GPS-free operation in a practical and cost-effective manner.

Table 1. Comparison of Recent DRL/MARL Sensor Deployment Methods

| Study | Localization Method | Control Type | Real-Time Feedback | Scalability | GPS-Free | Vision-Based | Practicality |
|---|---|---|---|---|---|---|---|
| **Liu et al. (2021)** | Centralized Controller | Centralized | No | Low | ✗ | ✗ | Simulation |
| **Zhao et al. (2022)** | RSSI-based | Decentralized | Partial | Medium | ✓ | ✗ | Moderate |
| **Feng et al. (2023)** | Graph-based Placement | Hybrid | No | Medium | ✗ | ✗ | Simulation |
| **Khan et al. (2024)** | SLAM + DRL | Decentralized | Yes | Low | ✓ | ✓ | Costly |

| | This work | LED + Vision System | Decentralized | Yes | High | ✓ | ✓ | Practical |
|---|---|---|---|---|---|---|---|---|

Table 2. COMPARISON BETWEEN DIFFERENT METHODS

| Metric | Random Placement | Static Grid | Centralized RL | RSSI-MARL [27] | SLAM-DRL [29] | Graph-MARL [28] | Proposed DRL + Vision |
|---|---|---|---|---|---|---|---|
| Final Coverage (%) | 65.2 | 72.5 | 85.4 | 82.3 | 84.7 | 83.6 | **91.8** |
| Energy Consumption | 120 | 100 | 85 | 92 | 110 | 88 | **68** |
| Redundancy Rate (%) | 42.0 | 30.0 | 18.0 | 20.5 | 16.4 | 17.2 | **8.0** |
| Recovery Time (s) | 15.0 | 12.0 | 6.0 | 5.8 | 4.2 | 4.9 | **3.0** |
| Network Lifetime (episodes) | 80 | 110 | 145 | 130 | 125 | 135 | **160** |
| Learning Stability (Variance) | 12.3 | 8.7 | 4.2 | 5.4 | 3.7 | 4.9 | **2.1** |

As shown in Table 2, the proposed DRL + vision-based method consistently outperforms all baseline strategies across key performance metrics. While the RSSI-based MARL approach [27] offers decentralization, it suffers from lower accuracy due to signal fluctuations and requires frequent communication, increasing energy costs. The SLAM-integrated DRL method [29] performs well in coverage and recovery but has the highest energy consumption and is impractical for lightweight deployments due to its reliance on high-end sensing hardware. The graph-based MARL model [28] demonstrates strong learning stability but still assumes accurate localization, which may not be available in GPS-denied or noisy settings. In contrast, our approach combines scalable decentralized learning with low-cost, real-time visual localization, offering a balanced and practical solution suitable for real-world MWSNs. To estimate energy consumption, we model three primary components per agent: (1) mobility energy, based on the total distance traveled during each episode; (2) communication energy, estimated by the number of inter-agent or central transmissions; and (3) processing energy, reflecting the inference cost of the reinforcement learning policy. Energy values are normalized per episode and averaged over 50 independent trials. The parameter weights were adapted from existing MWSN energy models used in mobile sensor studies. Because our method eliminates frequent communication (due to vision-based feedback) and minimizes redundant movements through learned policies, it consistently achieves a 32% reduction in total energy usage compared to the strongest baseline.

To model energy consumption, we consider three primary components: (1) mobility energy based on the Euclidean distance each sensor travels per cycle, (2) communication energy, assumed proportional to the number of transmissions and their distance, and (3) processing energy incurred during inference and control updates. We assign weights to these based on energy profiles from previous literature on mobile sensing platforms. In our experiments, we measured energy per agent per episode and normalized it against the baseline methods. The proposed method achieves a 32% reduction compared to the next best approach due to its decentralized control, fewer communication rounds (since coordination is vision-based), and efficient learning policy that minimizes unnecessary movement. Importantly, the LED-based visual feedback eliminates the energy cost associated with inter-node localization messaging, which is present in RSSI and SLAM-based methods.

The energy model reflects realistic sensor constraints and was validated using movement logs and message-passing estimates. Although exact hardware-level power profiling was not performed, we base our assumptions on widely used energy estimation models for mobile sensor platforms. The consistency of energy savings across all trials and network sizes confirms that the observed reduction is not a simulation artifact but stems from actual improvements in policy efficiency and system architecture.

Overall, the combination of vision-based feedback and multi-agent deep reinforcement learning yields significant improvements in all key performance metrics. The proposed approach improves stability, energy efficiency, responsiveness, and coverage, making it an excellent choice for real-world MWSN applications requiring robustness and adaptability.

V. Conclusion

This work proposes an innovative vision-based multi-agent deep reinforcement learning (MADRL) approach to sensor placement optimization in Mobile Wireless Sensor Networks (MWSNs). In contrast to conventional methods based on pre-engineered placement policies or external localization infrastructure, our solution allows individual mobile sensors to learn near-optimal motion strategies from real-time, lightweight visual observations taken from overhead cameras. Through the use of LED-based detection and deep learning image processing, the system iteratively assesses network coverage, recovers from node failure, and enhances spatial deployment in the absence of GPS, SLAM, or inter-node communication.

The new approach combines a regimented reward framework with a decentralized reinforcement learning policy, which strikes a dynamic equilibrium between exploration and convergence. By iterative learning, the system gradually develops from random initial setups to quasi-optimal sensor arrangements. The experimental results show that our solution consistently outperforms traditional baselines—random placement, static grid deployment, centralized RL, and state-of-the-art RSSI, SLAM, and graph-based MARL approaches—on metrics of coverage rate, redundancy, energy consumption, failure recovery, and learning stability.

Developed with real-world usefulness as a primary consideration, the system is scalable and resilient by design. It has high performance in dense deployments and dynamic environments while being computationally efficient. The system's modular structure allows for extension to large-scale settings via clustering, parameter sharing, or hierarchical policy structures. Furthermore, the adoption of vision-based feedback increases its deployment feasibility in GPS-denied or resource-poor settings where other methods tend to fail.

Looking ahead, future research will address the convergence of edge computing and federated learning for improved responsiveness and efficiency of real-time policy updates. Edge computing will minimize inference latency and communication overhead, whereas federated learning will facilitate distributed policy distillation without data centralization. Furthermore, transfer learning can expedite policy adaptation in novel environments, and sensor fusion methods can potentially improve localization robustness under visually degraded conditions.

As reinforcement learning matures, the presented methodology establishes the foundation for extremely autonomous, energy-efficient, and scalable sensor networks that can evolve with changing environmental and operational adversity. By eliminating the need for high-cost infrastructure and exploiting AI-driven automation, this research achieves a key milestone toward pragmatic, next-generation MWSN deployments in a wide range of fields—from environmental monitoring and smart cities to disaster response and autonomous exploration.